\documentclass{emulateapj}
\journalinfo{ApJ, in press}

\usepackage{graphicx}
\shorttitle{Outflow velocities in plumes  }

\shortauthors{Fu et al.}

\usepackage{multirow}

\begin{document}

\title{Measurements of outflow velocities in on-disk plumes from EIS/Hinode observations}

\author{Hui Fu\altaffilmark{1},  Lidong Xia\altaffilmark{1*},
Bo Li\altaffilmark{1}, Zhenghua Huang\altaffilmark{1}, Fangran
Jiao\altaffilmark{1}, Chaozhou Mou\altaffilmark{1}}

\altaffiltext{1}{Shandong Provincial Key Laboratory of Optical
Astronomy and Solar-Terrestrial Environment, School of Space
Science and Physics, Shandong University, Weihai 264209, China}
\altaffiltext{*}{xld@sdu.edu.cn}

\begin{abstract}
The contribution of plumes to the solar wind has been subject to
hot debate in the past decades. The EUV Imaging Spectrometer (EIS)
on board Hinode provides a unique means to deduce outflow
velocities at coronal heights via direct Doppler shift
measurements of coronal emission lines. Such direct Doppler shift
measurements were not possible with previous spectrometers. We
measure the outflow velocity at coronal heights in several on-disk
long-duration plumes, which are located in coronal holes and show
significant blue shifts throughout the entire observational
period. In one case, a plume is measured 4 hours apart. The
deduced outflow velocities are consistent, suggesting that the
flows are quasi-steady. Furthermore, we provide an outflow
velocity profile along the plumes, finding that the velocity
corrected for the line-of-sight effect can reach 10 km\,s$^{-1}$
at 1.02 $R_{\odot}$, 15 km\,s$^{-1}$ at 1.03 $R_{\odot}$, and 25
km\,s$^{-1}$ at 1.05 $R_{\odot}$. This clear signature of steady
acceleration, combined with the fact that there is no significant
blue shift at the base of plumes, provides an important constraint
on plume models. At the height of 1.03 $R_{\odot}$, EIS also
deduced a density of 1.3$\times10^{8}$ cm$^{-3}$, resulting in a
proton flux of about 4.2$\times10^9$ cm$^{-2}$s$^{-1}$ scaled to
1AU, which is an order of magnitude higher than the proton input
to a typical solar wind if a radial expansion is assumed. This
suggests that, coronal hole plumes may be an important source of
the solar wind.
\end{abstract}

\keywords{Sun: plume --- Sun: Doppler shift --- Sun: solar wind
--- EUV}

\section{Introduction}
 {  Solar plumes are ray-like bright structures extending
from the base of the corona into the high corona
(\citealp{DeForest2001b}; see also \citealp{Wilhelm2011} for a
recent review)}. They are the most conspicuous structures observed
during a total eclipse \citep{Hulst1950a,Hulst1950b}. They can be
observed in both visible light and EUV \citep[][and references
therein]{Wilhelm2011}. By tracing plume structures up to 15
$R_{\odot}$, \citet{DeForest1997,DeForest2001a} have demonstrated
that the structures seen in visible light and EUV are different
parts of the same phenomenon. Plumes are found to be rooted mainly
at {mid- and high-}latitudes where the global magnetic fields are
open into interplanetary space. Regarding magnetic field at their
foot-points, many people believe that plumes are associated with
mixed-polarity magnetic features with small bipolar fields
interacting with background unipolar fields
\citep{Wang1995,Wang2008}. So far, most studies on plumes are
carried out in and above polar coronal holes (PCHs) thanks to the
faint background that makes plumes easy to be observed. However,
some efforts have also been made to study the  { 
{properties}} of plumes at lower latitudes seen in
EUV~\citep{Wang1995,Zanna1999,Zanna2003,Wang2008,Tian2011}, white
light~\citep{Yang2011} and radio wavelengths~\citep{Woo1996}.
These studies have demonstrated the existence of plumes at lower
latitudes and have shown that the properties of mid- and
low-latitude plumes are similar to those of polar plumes
\citep{Zanna2003,Wang2008,Tian2011,Yang2011}.

 {  Coronal holes (CHs) are well believed to be the source
regions of the fast solar wind
\citep{Krieger1973,Zirker1977,Gosling1999,Hassler1999,Xia2003a}.
\citet{Krieger1973} were the first to associate a fast stream with
a coronal hole, while \citet{Zirker1977} suggested that all
coronal holes are connected to the fast solar wind.} Since plumes
are mainly found in CHs, their relationship with {the} solar wind
became a significant problem. Using the photometric data obtained
in the wavelength range between 171 \AA\ and 175 \AA\, combined
with a simple, semi-empirical model of {the} solar wind,
\citet{Walker1993} suggested that plumes are a possible source
region of the solar wind. By using {the} Doppler Dimming
technique, \citet{Gabriel2003,Gabriel2005} concluded that plumes
could provide about half of the mass that the solar wind carries
into interplanetary space. More recently, using high resolution
AIA/SDO observations, \citet{McIntosh2010} and \citet{Tian2011}
suggested that the high speed structures in plumes could provide
mass flux to the solar wind efficiently. Nevertheless, the
conclusion is still under debate. For instance, many studies
showed no clear evidence that plumes could provide sufficient
momentum and mass to solar wind flows
\citep{Wang1994,Habbal1995,Wilhelm2000,Hassler1999,Giordano2000,Patsourakos2000,Teriaca2003}
.

The primary reason for ruling out plumes as a solar wind source is
that the calculated mass flux in plumes is lower than needed by a
typical solar wind. Evidently, such a calculation requires that
the electron density and flow velocity be measured in plumes.
\citet{Habbal1995} calculated the electron density in both plume
and inter-plume regions in a PCH using the polarized white-light
observations. Combining with a two-fluid model, they obtained the
velocity {in} plume and inter-plume regions at different
altitudes, and found that the velocity of the inter-plume region
was faster than that of the plume region. Through analyzing {the}
O~{\sc{vi}} spectral data recorded by SUMER/SOHO,
\citet{Wilhelm1998} measured the velocity in plume{s} below
1.2~$R_{\odot}$, and found that the outflow velocity was smaller
than 18 km\,s$^{-1}$ after the line-of-sight (LOS) projection was
taken into account. In the same study, they also measured the
velocity in inter-plume regions with the Mg~{\sc{ix}} line, and
found that the LOS velocity was up to 34 km\,s$^{-1}$. Therefore,
they argue that the inter-plume lanes are the genuine source
regions of the fast solar wind. Through analyzing two images
simultaneously taken by the two EUVI telescopes on STEREO-A and
-B, \citet{Feng2009} reconstructed the location and inclination of
polar plumes. Then, the outflow velocity corrected {for} the LOS
effect is as small as 10 km\,s$^{-1}$ seen in O~{\sc{vi}} observed
by SUMER. They concluded that the plumes are unlikely a main
source of the fast solar wind when assuming a filling factor of
plumes to be 0.1 in CHs \citep{Ahmad1977}. Tracing the dynamic
features from 1.1 $R_{\odot}$ to 1.3 $R_{\odot}$ in white light
observations during the total eclipse, \citet{Belik2013} obtained
a propagation speed ranging from 32 km\,s$^{-1}$ to 146
km\,s$^{-1}$, with an average value of 67 km\,s$^{-1}$ in plume
structures.

The Doppler Dimming technique \citep{Rompolt1967} is often used to
derive the outflow velocity of off-limb plumes. In order to
investigate how outflow velocities change with height {at
altitudes below} 2~$ R_{\odot}$, \citet{Teriaca2003} explored
observations with SUMER/SOHO and UVCS/SOHO in a PCH, and from
their results, they concluded that {plumes are more or less
static, so inter-plume regions are suggested as sources of the
fast wind}. In contrast, \citet{Corti1997} found that in the
heliocentric range between 1.5 and 2.3~$R_{\odot}$, the outflow
velocity in plumes is about the same as that in inter-plume
regions. Using the same method, \citet{Gabriel2003,Gabriel2005}
derived an even faster outflow velocity in plumes than in
inter-plume regions below 1.6 $R_{\odot}$. Nevertheless, one word
of caution is that the Doppler dimming method strongly depends on
empirical assumptions of the electron density and ion temperature,
thus may lead to different results with different assumptions
\citep{Wilhelm2011}.

Obviously, measuring the Doppler shift on the disk is a more
straightforward way of getting the outflow velocity in plumes.
Some spectroscopic studies have been made for on-disk plumes. By
analyzing the Ne~{\sc{viii}} 770 \AA\ line observed in a north PCH
with SUMER, some authors \citep{Hassler1999,
Hassler2000,Wilhelm2000} found that the base of the plumes does
not show a clear blue shift compared to other CH regions. However,
the above studies are based on observations with SUMER, which has
limited spectral lines at coronal temperatures, so the measured
radiation is mainly emitted by the bright cores at the bottom of
plumes.

In this study, we use spectroscopic observations with the EUV
Imaging Spectrometer (EIS) \citep{Culhane2007} on board the Hinode
satellite \citep{Kosugi2007}, which includes many coronal lines
and covers a wide temperature range. It gives us an opportunity to
study the Doppler velocities in the extended structure of plumes.
Meanwhile, the influence of background is lower when the extended
structure of plumes projects on the dark background of CHs. This
offers a bigger advantage in observing plumes than SUMER does. The
paper is organized as follows. In Section 2, we describe the
observations and data analysis methods. The results are shown in
Section 3. The significance of our results associated with the
origin of the solar wind is discussed in Section 4. A conclusion
is presented in Section 5.

\section{Observation and data analysis}\label{sec2}

Four data sets obtained by EIS are used in the present study.
Among them, one data set was recorded in a mid-latitude region,
the others were recorded in the north pole. Because we are
interested in plumes that project on CHs, the data sets with
long-exposure time and 2$^{''}$ wide slit are chosen in order to
have a good signal-to-noise ratio. The details of the scanning
period, observed location, size of field of view (FOV) and
exposure time for the selected data sets are listed in Table 1. In
Table 2, we list eight lines with formation temperatures ranging
from 50000 K to 2 million K (MK) that are used for analyzing the
intensity contrast of plumes and the CH background. Among them,
five lines with good signal-to-noise ratio and no obvious blends
are chosen to deduce the Doppler shift. According to
\citet{Young2007} and \citet{Tian2010}, these lines can be well
fitted by a single Gaussian. It is noteworthy that the data taken
in 2007 have been used by \citet{Banerjee2009} to search for
signatures of Alfv\'{e}n waves and \citet{Tian2010} to describe
the scenario of the nascent fast solar wind guided by expanding
magnetic funnels. The difference between the study of
\citet{Tian2010} and our work will be further discussed in
Section~\ref{sec3}. We also use the data from EIT on board SOHO,
and AIA on board SDO to identify plumes.

The EIS data are calibrated by standard procedures including
dark-current, flat-field, removal of cosmic rays, hot and warm
pixels, and radiometric calibration via the routine
\emph{eis\_prep.pro} provided by the instrumental team. EIS has
two CCDs that cover the wavelengths from 170 \AA\ to 210 \AA\, and
250 \AA\ to 290 \AA\, respectively. Images taken by the two CCDs
have offsets along the slit direction and raster-scanning
direction due to the arrangement of the CCDs. According to our
study, an offset of 16$^{''}$ was found between the two CCDs along
the slit direction, which is in agreement with the results found
by \citet{Tian2010} and \citet{Demoulin2013}. The procedure
\emph{eis\_wave\_corr.pro} in Solar software (SSW) is used to
compensate for the orbital variations of the EIS line centers
\citep{Kamio2010} and the slit tilt caused by tilt of slits
relative to the axes of the EIS CCDs.

Wavelength calibration is an important step for obtaining precise
Doppler shifts of spectral lines. Unfortunately, EIS has neither
absolute wavelength calibration on board nor cold lines as
reference. However, there are two methods for obtaining the
absolute Doppler shift, one assuming that the LOS velocity of the
off-limb plasma is zero \citep{Peter1999,Kamio2007}, the other
assuming that the quiet-Sun (QS) region is almost quiet and can be
used as a reference. The data set taken in 2007 includes both
off-limb and QS observations, which provides an opportunity to
verify the two methods. Because only Fe~{\sc{x}} 184.54 \AA,
Fe~{\sc{xii}} 195.12 \AA\ and Fe~{\sc{xiii}} 202.04 \AA\ have
reliable signal-to-noise ratio in the off-limb data, we use them
to derive the required referent wavelengths. We first obtain a
spectrum by averaging all the spectra from $65''$ to $143''$ above
the limb \citep{Tian2010}. By assuming that the average off-limb
spectrum is zero-shifted, we then calculate the line shifts of an
average spectrum of a QS region, {where the Doppler velocities are
found to be -3.0$\pm$2.8 km\,s$^{-1}$, -3.3$\pm$2.8 km\,s$^{-1}$,
-3.9$\pm$2.8 km\,s$^{-1}$ (here only fitting errors are
considered)} in Fe~{\sc{x}} 184.54 \AA, Fe ~{\sc{xii}} 195.12 \AA\
and Fe~{\sc{xiii}} 202.04 \AA, respectively, which are consistent
with the results found by \citet{Peter1999} and
\citet{Dadashi2011}. By using cospatial and nearly cotemporal
SUMER and EIS data, \citet{Dadashi2011} obtained  {  blue
shifts (upward motions) between -2 to -4 km\,s$^{-1}$} at
temperatures between 1 and 1.8 MK. Therefore, we select the line
center position averaged over a QS region, where a blue shift of
3.5 km\,s$^{-1}$ is assumed, as the reference to calculate the
absolute velocities of plumes in CHs.

 {  An error analysis is necessary but not straightforward.
Three components of errors contribute to the total error of the
derived Doppler velocities: 1) the fitting error
$\sigma_{\mathrm{fitting}}$, 2) the error associated with
assigning an absolute speed (-3.5 km\,s$^{-1}$ in this paper) to
the QS region $\sigma_{\mathrm{abs}}$, and 3) the error associated
with the unphysical noisy stripes of the derived Dopplergrams
$\sigma_{\mathrm{noise}}$. It turns out that
$\sigma_{\mathrm{fitting}}$ is the smallest among the three
(varying between 0.1 and 0.2 km\,s$^{-1}$), for the spectra
employed as the input to the fitting procedure are obtained by
averaging the relevant spectra over a substantial area and
therefore are close to a perfect Gaussian. As for
$\sigma_{\mathrm{abs}}$, \citet{Dadashi2011} established that for
spectral lines with formation temperatures between 1 and 1.8 MK,
this error ranges from 2.0 to 3.0 km\,s$^{-1}$ (see their Table
3), here the median value of 2.2 km\,s$^{-1}$ is chosen for our
purpose. $\sigma_{\mathrm{noise}}$ can be estimated by computing
the standard deviation of the derived Dopplergrams over a selected
area with visible vertical and/or horizontal artifacts. For
dataset 1, the off limb region is chosen whereby
$\sigma_{\mathrm{noise}}$ is estimated to be 1.7 km\,s$^{-1}$.
However, for datasets 2, 3, 4, the Dopplergrams are not as well
organized as the one for dataset 1 (compare Figures 5(b), 7(b1),
7(b2) with Figure 3(d2)): some horizontal and vertical stripes
show up in e.g., the middle part in the lower half of Figures 5(b)
and 7(b2)), their origin likely to be unphysical. Therefore we
compute $\sigma_{\mathrm{noise}}$ in the regions where these
artifacts are most evident in order not to underestimate this
error, finding that $\sigma_{\mathrm{noise}}$ reads 1.8 (2.1, 3.3)
km\,s$^{-1}$ in dataset 2 (3, 4). We then evaluate the desired
total error $\sigma$ according to $\sigma =
\sqrt{\sigma_{\mathrm{fitting}}^2+\sigma_{\mathrm{abs}}^2+\sigma_{\mathrm{noise}}^2}$,
the results being 2.8, 2.8, 3.0, and 4.0 km\,s$^{-1}$ for datasets
1 to 4, respectively.}

The identification of plumes in on-disk images is not
straightforward. To make plumes easier to identify, a method
called Multi-scale Gaussian Normalization (MGN)~\citep{Morgan2014}
is applied. This method helps reveal fine scales without the loss
of larger-scale information.

\section{Results}\label{sec3}

In Figure 1, we show the intensity map (middle panel) derived from
the EIS raster scan in Fe~{\sc{xiii}} 202.04~\AA\ on 2007 October
10, together with the corresponding EIT 195~\AA\ image (left
panel) with a larger field of view (FOV) and processed image of
intensity map using the method MGN. As previous studies have
demonstrated, plumes are identified as faint and diffuse blobs of
emission-enhanced structure with a brighter core seen in EUV on
the disk \citep{Wang2008}. Accordingly, we may identify a number
of plumes in this data set. Among them three plumes outlined by
the green solid lines (labeled PL1, PL2 and PL3) have clearly
extended structures in the EIS FOV by an inspection of the
corresponding processed image. The three plumes are extending to
heights of around 80 to 100 arcsec, which are further divided into
three sub-regions as bottom, middle, and top segments marked by
the white rectangles for further analysis. Please note that plumes
are not clearly seen in the EIT image, perhaps due to the lower
sensitivity of EIT relative to EIS. Using the same procedures, we
also identify and select three plumes in the other three data sets
for analysis.

In CH plumes, the measured emission mainly comprises two
components: the plumes themselves and CH background. In order to
estimate their specific contribution, here we define a parameter
called the \emph{relative radiation intensity}
(\emph{RRI=intensity of selected region/mean QS intensity}).
\emph{RRI} represents the intensity ratio of a selected region and
the mean QS region. Fortunately, the EIS FOVs cover QS regions in
every event.

In Figure 2, we display the \emph{RRI} parameter of three plumes
measured in the spectral lines of interest recorded on 2007
October 10. For comparison, the \emph{RRI} measurement of a
typical dark region as seen in hot coronal lines in the CH is also
presented. From Figure 2, one finds that the intensity of
He~{\sc{ii}} is generally weaker in the entire CH region including
plumes than in the QS region, which is consistent with the
previous EIT observations. For the dark CH region, the \emph{RRI}
has a value larger than 1 in both Fe~{\sc{viii}} and Si~{\sc{xii}}
lines with formation temperatures below 1 MK. This may be caused
by the existence of the underlying bright network structures (also
see Figure 3). The \emph{RRI} decreases quickly to 0.2 in
Fe~{\sc{x}}. This is a typical result observed previously by SUMER
\citep{Xia2003b}. However, for the coronal lines whose formation
temperatures exceed 1 MK, the \emph{RRI}s of the dark CH region
are all less than 0.1 and nearly constant (0.09, 0.08, 0.08 in
Fe~{\sc{xii}} 195.12 \AA, Fe~{\sc{xiii}} 202.04 \AA\ and
Fe~{\sc{xv}} 284.16 \AA, respectively). This is very likely caused
by the stray light and the reason why we can only obtain smaller
blue shifts from these three lines in the dark region, which are
comparable with those obtained in the QS region (see further
discussion below).

For plumes, the \emph{RRI} reaches maximum in the extended
structure for Fe~{\sc{viii}} and Si~{\sc{xii}} lines, which can be
larger or smaller than 1. These lines seem to be mainly emitted by
the network structures underlying the extended structure of
plumes. Nevertheless, the \emph{RRI} reaches a minimum for the
Fe~{\sc{xii}} and Fe~{\sc{xiii}} lines and then increases again
with formation temperature.  {{  Moreover, line intensities
of Fe~{\sc{xii}}, Fe~{\sc{xiii}}, Fe~{\sc{xv}} are much stronger
than those of the CH background as shown in Figure 2.
Quantitatively, this may be illustrated by examining the specific
\emph{RRI} values. Take the plume PL3, the weakest in emission
among the three plumes, for instance. For its top segment, the
\emph{RRI} reads 0.26 in Fe~{\sc{xii}}, 0.30 in Fe~{\sc{xiii}},
and 0.34 in Fe~{\sc{xv}}. As a comparision, the background coronal
hole corresponds to \emph{RRI} values of 0.09, 0.08, and 0.08 for
the three passbands, respectively. Consequently, the contribution
of the plume plasma to the overall emission can be estimated to be
65\% in Fe~{\sc{xii}}, 73\% in Fe~{\sc{xiii}}, and 76\% in
Fe~{\sc{xv}}. Given that the top segments are subject to a
stronger contamination from the emission from the background
coronal hole, and that the plume PL3 suffers from the strongest
contamination among the three plumes, we then deduce that in
Fe~{\sc{xii}} (Fe~{\sc{xiii}}, Fe~{\sc{xv}}), at least 65\% (73\%,
76\%) of the emission comes from the plume plasma itself.
Actually, the derived relative contribution is consistent with
those obtained by \citet{Tian2011}.}}

{   { Given that the plumes were deduced to be less hot than
inter-plume regions \citep{DeForest1997, Hassler1997,
Banerjee2000}, one may question why they emit in a line as hot as
Fe XV as well. \citet{Raouafi2008} find plumes are associated with
hot and transit X-ray jets, and the lifetimes of those jets range
from minutes to a few tens of minutes. It then follows that the
plumes we examine may be also associated with X-ray emitting, hot
structures as well. However, a close examination indicates that
the main structures of plumes are almost quasi-steady. The plumes
in data sets 2 to 4 have lifespans exceeding 4 hours, and their
morphology remains stable in the time interval we examined (see
the AIA movies and running difference movies: movie20100904,
movie20100904diff, movie20110111 and movie20110111diff in the
supplement). In AIA movies (especially for data set 20110111),
like \citet{Raouafi2008} found, there are many small transient
structures at the base of plumes. Those explosive events are
smaller in size (only several arcsecs) and weaker in intensity
(the intensity changes by no more than 10\%) compared to
quasi-steady plume structures. For comparison, there is a classic
jet in AIA movie20100904 ($t$=00:10-00:25, $x$=-75$''$,
$y$=900$''$) for which the intensity increased by 35\% in the 171
band, and by 50\% in the 193 band when the jet erupts. We suspect
that the major plume plasmas can be supplied by those small
transient events. As for the plumes in data set 1 acquired in
2007, the EIT 195 movie suggests that they are stable and not
associated with big explosive events. One then naturally ask
whether stable plumes also emit in hot lines. Actually,
\citet{DeForest1997} also noticed that plumes are seen not only in
EIT 171 movie and EIT 195 images, but also in the Fe~{\sc{xv}} 284
band. One possibility is that, although the temperature of plumes
is lower than in inter-plume regions, but plumes themselves may
contribute to a significant fraction of the corresponding emission
due to their considerably higher densities than the surrounding
coronal hole plasmas. Another possibility that cannot be ruled out
is that the plume structures are composed of hot sub-resolution
transients. Not resolved adequately in current observations, these
transients may make the plume structures appear quasi-stable.}}

The intensity and Dopplergram maps derived from the
Fe~{\sc{viii}}, Si~{\sc{vii}}, Fe~{\sc{x}}, Fe~{\sc{xii}} and
Fe~{\sc{xiii}} lines are displayed in Figure 3. Figure 3 brings to
our attention that the strong blue-shift patches derived from hot
lines of Fe~{\sc{x}}, Fe~{\sc{xii}} and Fe~{\sc{xiii}} coincide
with the extended structures of the identified plumes. In
contrast, this coincidence is not clear in cooler lines of
Fe~{\sc{viii}} and Si~{\sc{vii}}, where the intensity and
Dopper-shift patterns reflect mainly local network structures but
not plumes. As we have mentioned, there are smaller blue shifts in
the dark region of the CH, which is contoured by white lines in
Figure 1 (middle panel) and red lines in Figure 3. The Doppler
shift is -2.4 km\,s$^{-1}$ in Fe~{\sc{xii}} 195.12 \AA\ and -3.4
km\,s$^{-1}$ in Fe~{\sc{xiii}} 202.04 \AA\ respectively, which is
comparable with the QS region. Combining with the fact that in the
region where plume structure projects on the CH background the
radiation is mainly from plume structure for those lines with
formation temperatures in excess of 1 MK, we can draw an important
conclusion that the Doppler shift in the region where plume
structure projects on CH reflects the outflow in plume structures.

The above analysis has further demonstrated our identification of
the extended structure of plumes, which differs from the
explanation of \citet{Tian2010}, where they interpreted the
blue-shift patches in Figure 3 as evidence of the nascent fast
solar wind originating from the underlying magnetic networks. In a
later paper, \citet{Tian2011} have also clarified that in this
data set the blue-shift patches in CH are mainly contaminated by
the plumes rooted at the boundary of the PCH (see Figure 8 in
their paper).

Figure 4 shows the outflow velocity along the plumes derived from
the Fe~{\sc{viii}}, Si~{\sc{vii}}, Fe~{\sc{x}}, Fe~{\sc{xii}} and
Fe~{\sc{xiii}} lines. At first glance, it is a little strange that
the Doppler shifts seen in the very hot lines are much different
from those in colder lines at the same height. {   { Then,
which ones reflect the real Doppler shifts of plume structures?
Here we argue that the Doppler shifts of the hottest lines whose
formation temperatures exceed 1 MK reflect more faithfully the
real outflow in plumes. As shown in Figure 3(a1) and 3(b1), the
plume structures are almost invisible in the lines of
Fe~{\sc{viii}} and Si~{\sc{vii}}, which means that the emission
from the CH background dominates in these cool lines. This may be
further corroborated by Figures 3(a2) and 3(b2), where the
Dopplergrams are shown. In contrast to the Doppler shift features
in hot lines (Figures 3(c2-e2)), the Doppler features are not
organized in linear shapes as seen in the intensity diagrams. In
practice, we derive our outflow speeds for plumes using only the
Fe~{\sc{xiii}} line. The reason we do not use the hottest line,
Fe~{\sc{xv}}, is that the spectral shape is not as good as the
Fe~{\sc{xiii}} line.}} In this way, we can draw a further
conclusion that Doppler shifts increase with height in the
extended structure of plumes. Furthermore, assuming that the
plumes are extending radially, one can deduce the outflow velocity
of the plume plasma by correcting the latitude/LOS effect of the
plumes. For plumes of the 2007 October 10 data set, the latitude
of the plumes is about 60 degree.  {{  The outflow velocity
derived from the hottest line increases with height in the plumes,
and reads about $10\pm5.6$ km\,s$^{-1}$ at 1.02 $R_{\odot}$,
$15\pm5.6$ km\,s$^{-1}$ at 1.03 $R_{\odot}$, and $25\pm5.6$
km\,s$^{-1}$ at 1.05 $ R_{\odot}$.}}

If a plume is observed near the equator and its orientation is
nearly parallel to the local radial direction, it will allow us to
derive its outflow velocity directly from Doppler shifts with a
less strong projection effect. Fortunately, we observed a low
latitude plume in the data set of 2011 Jan 11 as show in Figure 5.
The plume (marked by the ellipse) is located at the boundary of a
big low latitude CH. Although it is not very clearly seen in the
EIS Fe XII 195.12 \AA\ intensity map, it can be easily identified
in the AIA 171 \AA\ movie. The Doppler shifts of the plume
structure and CH region (marked by the rectangle) are shown in
Figure 6. Green, blue and red curves represent the Doppler shifts
of QS, CH and plume regions, respectively. We find the Doppler
shift of plume structure is larger than 20 km\,s$^{-1}$ in the
coronal line Fe XII 195.12 \AA, which is consistent with the
corrected velocity derived from the 2007 Oct 10 event.

 { {A question may arise as to whether these blue-shift
patches are quasi-steady outflows or velocity patterns associated
with transient events. This is necessary given that
\citet{DeForest2001a} and \citet{Raouafi2008} have shown that
plumes appear to be recurrent and bursty. To address this
question, we follow a plume imaged by four hours apart on 2010 Sep
4, which are shown in Figure 7. The top panels show the intensity
and Dopplergram observed from 18:30 to 22:02, and the bottom
panels are for the measurements between 22:04 and 01:36 the
following day. Among several plume structures, one projected on CH
and marked by the white rectangle is chosen for further analysis.
From Figure 7, the shape of the plume does not change
substantially in the two scans. Now the question is whether the
plumes may have pulsed twice? We then examined the AIA movie for
the corresponding time interval (movie20100904, now supplemented
to Figure 7), and found that indeed the plume persisted for more
than 7 hours during these two EIS scans. From this we conclude
that the blue-shift patches reflect a quasi-steady outflow pattern
in the plume structure itself rather than being associated with
big transient jets.}}

\section{Discussion}\label{sec4}

In the past decades, many studies have been carried out in search
of the relationship between plumes and the solar wind
\citep{Walker1993,Wang1994,Habbal1995,Hassler1999,Wilhelm2000,Giordano2000,Patsourakos2000,Teriaca2003,Gabriel2003,Gabriel2005,McIntosh2010,Tian2011}.
One of the most important studies is the measurement of the
outflow velocity in plume and inter-plume regions. The outflow
velocity of off-limb plumes has been measured during the SOHO era
\citep{Wilhelm1998,Corti1997,Teriaca2003,Gabriel2003,
Gabriel2005}. However, these measurements lead to contradictory
results. On the disk, \citet{Hassler1999, Hassler2000,
Wilhelm2000} measured Doppler shifts at the bright footpoints of
plumes and found no obvious velocity signature since SUMER has
limited spectral lines at coronal temperatures. In this study, we
analysed several plumes observed by EIS on the disk. We derived
their outflow velocity from hot coronal lines. In polar and low
latitude regions alike, a quasi-steady outflow is present in the
plumes. The outflow velocity of plumes increases with height in
plumes and is found to be about 25 km\,s$^{-1}$ at 0.05
$R_{\odot}$ above the surface. In Figure 9, we present the outflow
velocity as a function of altitude measured by different authors,
and insert our results shown as red crosses. This may provide
further information on the velocity of the plume plasma from the
solar surface out to 1.4 $R_{\odot}$, which should serve as an
important constraint on plume models.

As mentioned in the Introduction, to estimate the possible
contribution of plumes to the solar wind, we need to know the
electron density to calculate the mass flux. At the height of 1.03
$R_{\odot}$, we also estimate the electron density by using the
CHIANTI database in line pair Fe~{\sc{xiii}} 202.04~\AA
/Fe~{\sc{xiii}} 203.82~\AA\ \citep{Young2009} and obtain an
electron density of 1.3$\times10^8$ cm$^{-3}$. Assuming all the
mass expanding from the plumes can travel to the Earth, it can
contribute a proton flux of 4.2$\times10^9$ cm$^{-2}$s$^{-1}$ at
1~AU, which is an order-of-magnitude higher than that of a typical
fast solar wind if a radial expansion is assumed. It should be
mentioned that the plumes we analysed are mainly rooted at the
boundary of CHs. The CH boundary is considered as a possible
source of the slow solar wind
\citep{Wang2009,Subramanian2010,Madjarska2012}. Yet, the large
mass flux that plumes can provide suggests that, plumes may be an
important source of the (fast and/or slow) solar wind, but they do
experience substantial lateral expansion and/or mass exchange with
neighboring inter-plume plasmas.

 { {In our study, we found that the CH regions do not show
more significant blue shifts than the QS regions, which seems to
contradict the consensus that CHs are the sources of the fast
solar wind. The reason for this apparent contradiction is two
fold. First, the radiation that EIS measured in the CHs is
dominated by the stray light from the surrounding QS and active
regions. Second, the solar wind may have been braked at the
studied altitudes before being accelerated at higher altitudes.}}
According to our analysis, we prefer the former interpretation. To
corroborate this, we have estimated the electron density in the QS
region and dark region of the CH observed on 2007 October 10, the
obtained density is about 2.2$\times10^8$ cm$^{-3}$ and
3.7$\times10^8$ cm$^{-3}$ respectively. However, it is almost
impossible that a CH region was denser than a typical QS region.
Combined with the fact that the \emph{RRI} parameter is almost
constant in lines with high temperatures found in Section 3, this
means that the radiation from the hot coronal lines is mainly from
the surrounding bright area most possibly scattered by the
instrument. A further study on the CH region, especially using
spectroscopic observations with very low stray-light spectrometer,
is needed to understand this contradiction.

\section{Conclusion}\label{sec5}

We have measured the outflow velocity at coronal heights in
several on-disk long-duration plumes observed by EIS on board
Hinode, which are located in coronal holes and show significant
blue shifts throughout the entire observational period. The
deduced outflow velocities are quasi-steady and can reach 10
km\,s$^{-1}$ at 1.02 $R_{\odot}$, 15 km\,s$^{-1}$ at 1.03
$R_{\odot}$, and 25 km\,s$^{-1}$ at 1.05 $R_{\odot}$ after
corrected for the LOS effect. This clear signature of steady
acceleration, combined with the fact that there is no clear
blueshift at the base of plumes, provides an important constraint
on plume models. At the height of 1.03 $R_{\odot}$, EIS also
deduced a density of 1.3$\times10^{8}$ cm$^{-3}$, resulting a
proton flux of 4.2$\times10^9$ cm$^{-2}$s$^{-1}$ scaled to 1AU,
which is an order of magnitude higher than necessary for the
proton input to a typical solar wind if a radial expansion is
assumed. This suggests that plumes may be an important source of
the solar wind.

\acknowledgements{The authors would like to thank the anonymous
referee for helpful comments on the manuscript. Hinode is a
Japanese mission developed and launched by ISAS/JAXA, with NAOJ as
domestic partner and NASA and STFC (UK) as international partners.
It is operated by these agencies in co-operation with ESA and NSC
(Norway). We are grateful to Shadia Rifai Habbal for helpful
discussions. This work was supported by grants NSBRSF
2012CB825601, NNSFC 41274178, 40904047, 41174154, 41274176, and
the Ministry of Education of China (20110131110058 and
NCET-11-0305). }

\clearpage

\begin{table}
\begin{center}
\caption{ EIS Observations Used in This Study.\label{tbl-1}}
\begin{tabular}{crrrrrrrrrrr}
\tableline\tableline
 Obs.ID & Scanning Period & Location & FOV(arcsec$^2$) & Slit
 & Exposure Time(s)\\
\tableline
1 &2007 Oct 10 14:03-18:16 &9,889 &202x512  &2$^{''}$ &150\\
2 &2010 Sep 04 18:30-22:02 &-11,889 &280x256 &2$^{''}$ &90\\
3 &2010 Sep 04 22:04-01:36 &-11,890 &280x256 &2$^{''}$ &90\\
4 &2011 Jan 11 09:45-13:17 &-23,384 &280x256 &2$^{''}$ &90\\
\tableline
\end{tabular}
\end{center}
\end{table}

\begin{table}
\begin{center}
\caption{Emission Lines Used in This Study.\label{tbl-2}}
\begin{tabular}{crrrrrr}
\tableline\tableline
 Line.ID &Ion & Wavelength(\AA) & Log(T/K) & Region$^{a}$ & Doppler shift\\
\tableline
1 &He~{\sc{ii}} &256.32 &4.7 &TR & No\\
2 &Fe~{\sc{viii}} &185.21 &5.6 &LC & Yes\\
3 &Si~{\sc{vii}} &275.35 &5.8 &LC & Yes\\
4 &Fe~{\sc{x}} &184.54 &6.0 &Cor & Yes\\
5 &Fe~{\sc{xii}} &195.12 &6.11 &Cor & Yes\\
6 &Fe~{\sc{xiii}} &202.04 &6.2 &Cor & Yes\\
7 &Fe~{\sc{xiv}} &274.20 &6.25 &Cor & No\\
8 &Fe~{\sc{xv}} &284.16 &6.3 &Cor & No\\
\tableline
\end{tabular}
\tablenotetext{a}{TR: Transition Region; LC: Low Coronal region;
Cor: Coronal region.}
\end{center}
\end{table}

\begin{figure}
\centering
\includegraphics[width=1.0\textwidth]{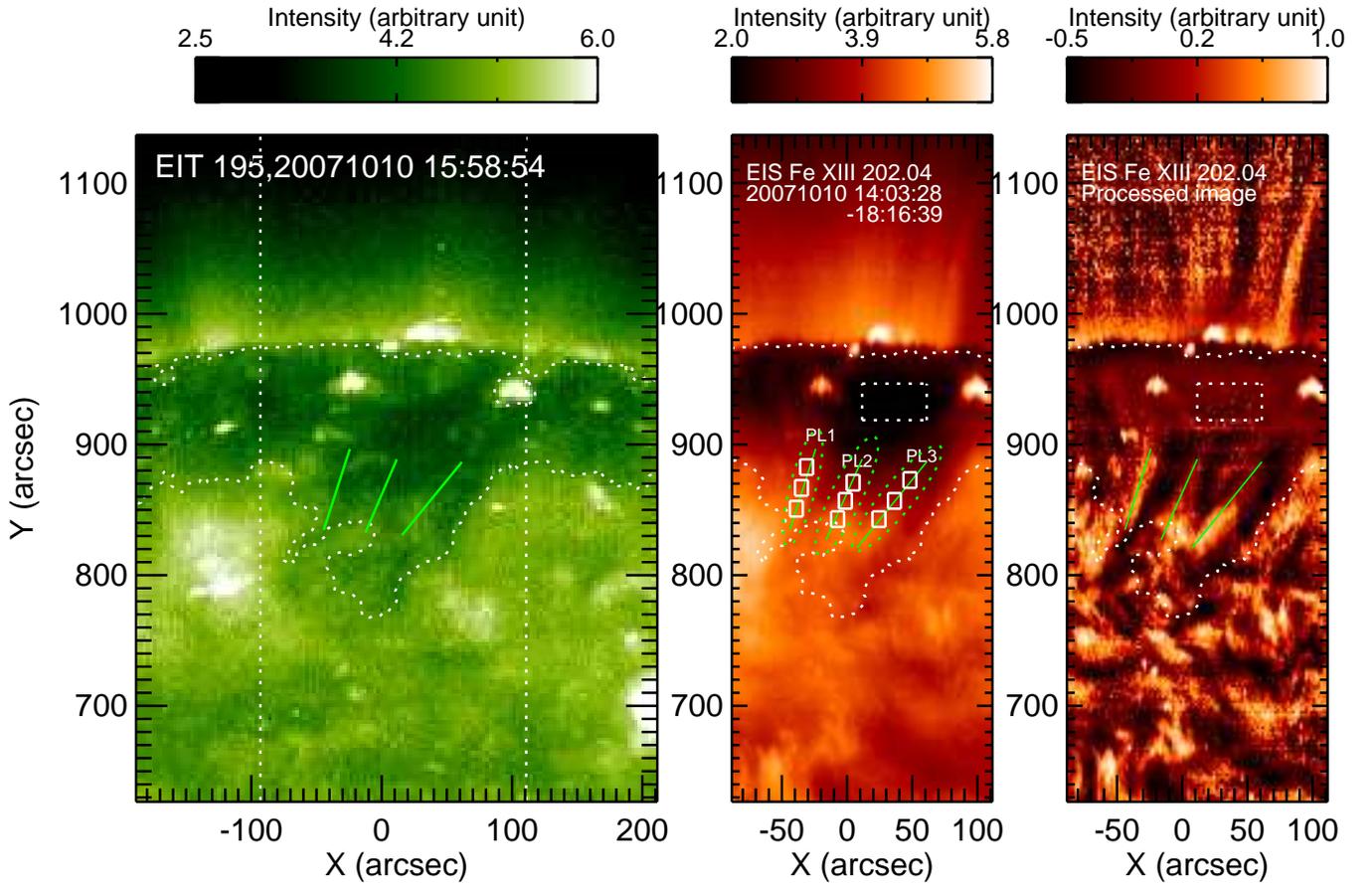}
\caption{Images of the north PCH observed on 2007 October 10. The
left panel shows the EIT 195 \AA\ image with EIS field of view
overplotted (vertical dotted lines). The middle panel shows the
intensity map recorded by EIS in Fe~{\sc{xiii}} 202.04 \AA, and
the right panel shows the processed image of intensity map of EIS.
The white contour lines outline the boundary of the CH, and the
small region marked by the dotted line rectangle represents the
typical dark region of the CH. The three tilted green full lines
represent the center of the plume structures which are better seen
in the processed image (right panel), and the three full line
rectangles in each plume represent the plume structure from bottom
to top.  {An EIT movie for the corresponding time interval is
available online (movie20071010.mpeg).}}
\end{figure}

\begin{figure}
\centering
\includegraphics[width=1.0\textwidth]{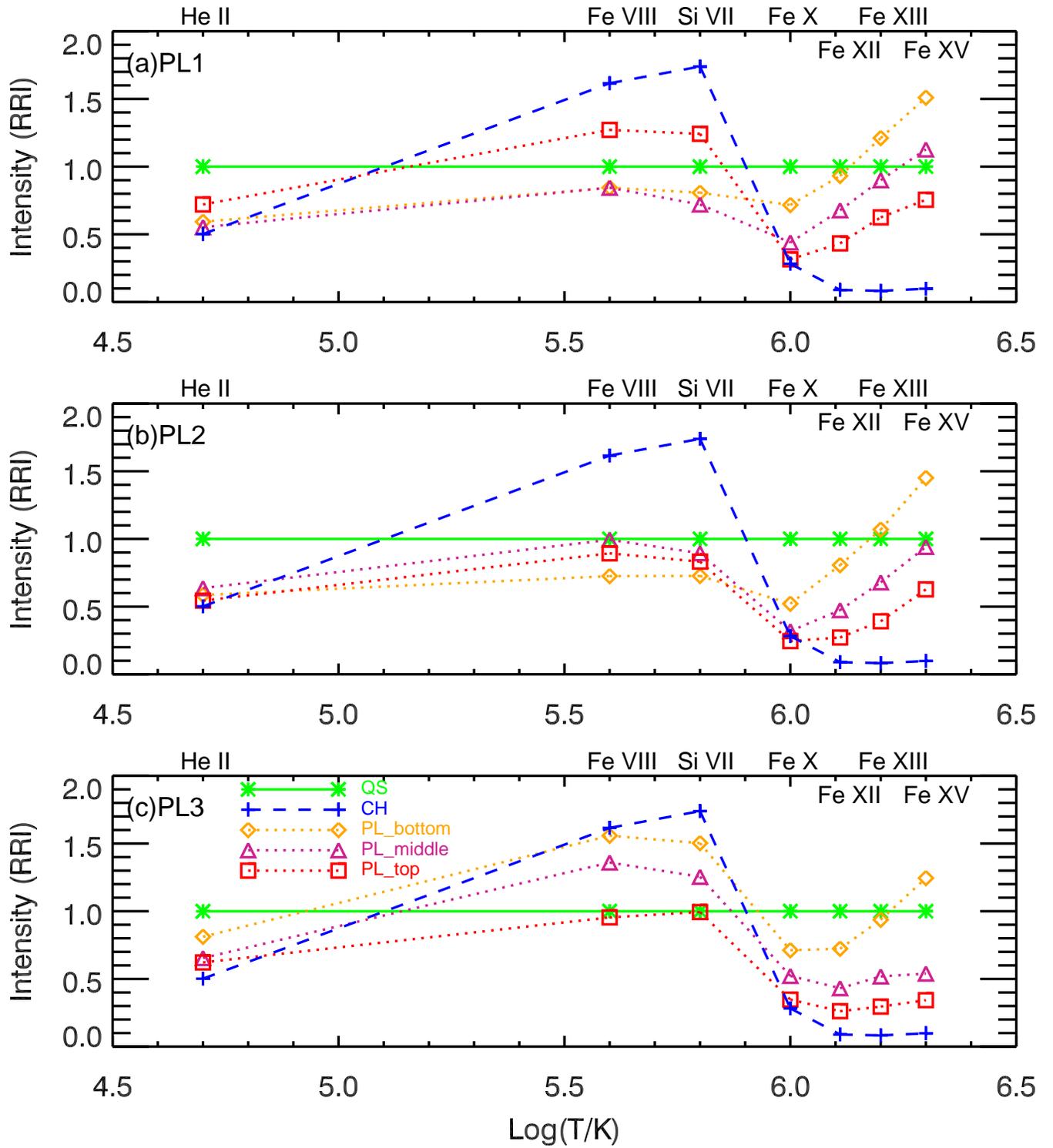}
\caption{The variation of \emph{RRI} with line formation
temperature in plumes and the dark CH region. Green and blue lines
represent the QS and CH regions, orange, violet red and red lines
represent the plume structure from bottom to top.}
\end{figure}

\begin{figure}
\centering
\includegraphics[width=1.0\textwidth]{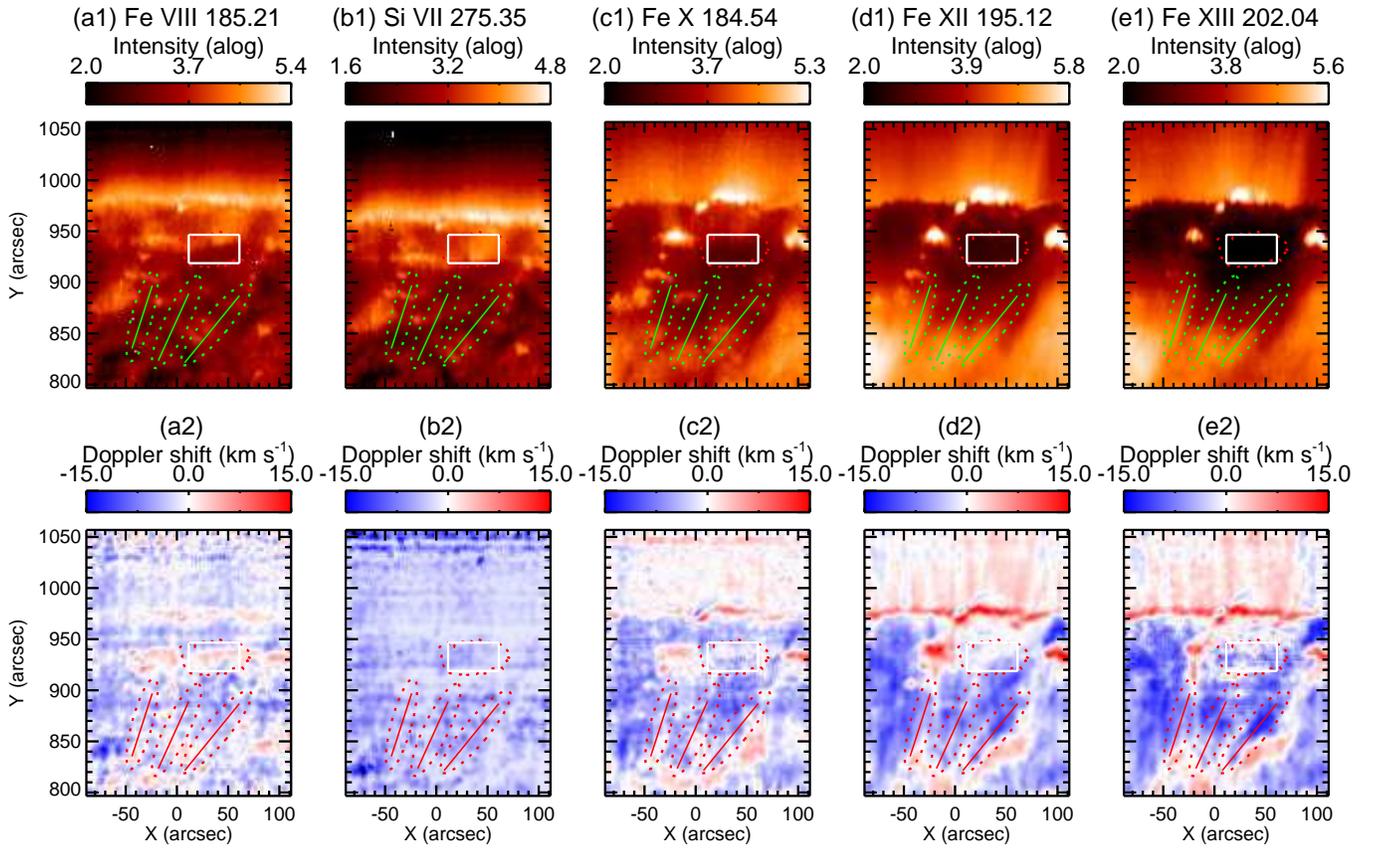}
\caption{The intensity map and Dopplergram as recorded in the
Fe~{\sc{viii}} 185.21 \AA, Si~{\sc{vii}} 275.35 \AA, Fe~{\sc{x}}
184.54 \AA, Fe~{\sc{xii}} 195.12 \AA\ and Fe~{\sc{xiii}} 202.04
\AA\ lines on 2007 Oct 10. The plume structures are clear in
Fe~{\sc{xii}} 195.12 \AA\ and Fe~{\sc{xiii}} 202.04 \AA, and
coincide with strong blue-shift patches derived from these two
lines.}
\end{figure}

\begin{figure}
\centering
\includegraphics[width=1.0\textwidth]{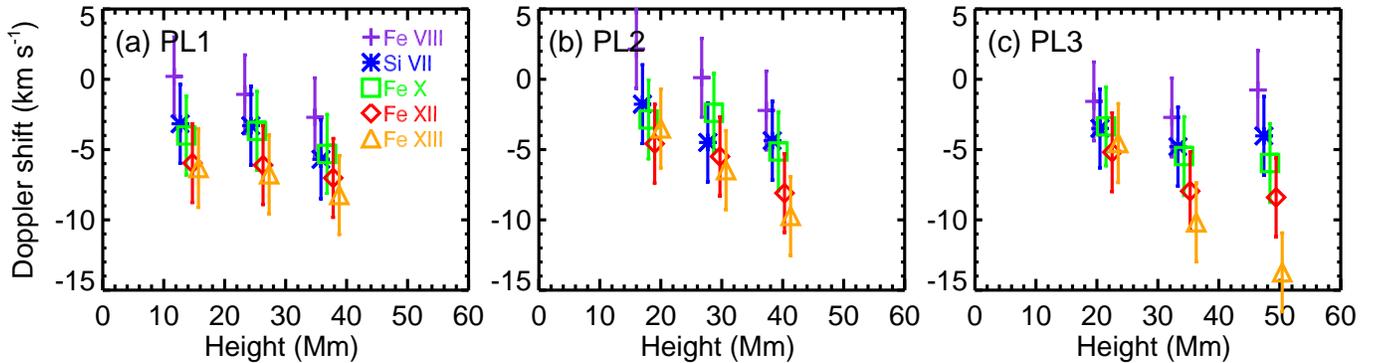}
\caption{The Doppler shifts along the extended structure of plumes
observed on 2007 Oct 10. For a description of how the errors are
found, see text.}
\end{figure}

\begin{figure}
\centering
\includegraphics[width=1.0\textwidth]{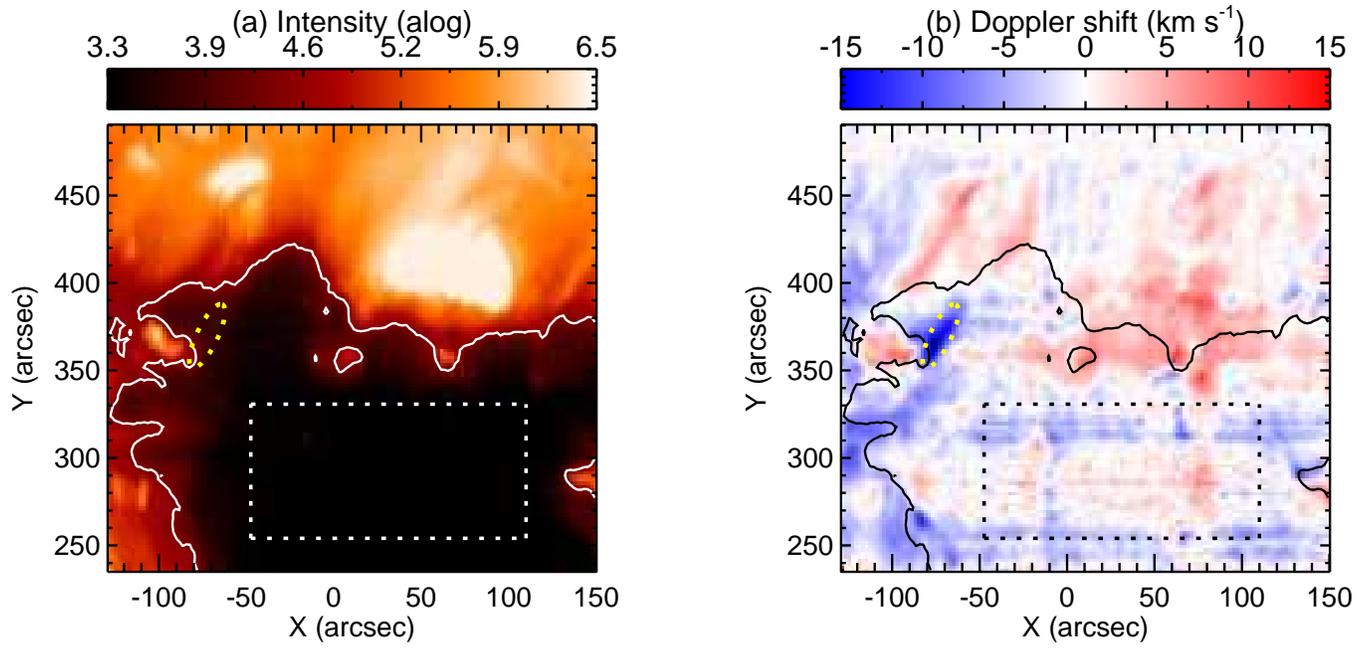}
\caption{The intensity map and Dopplergram as recorded by in the
Fe~{\sc{xii}} 195.12 \AA\ line on 2011 Jan 11. The plume is
outlined by yellow dotted ellipse and the rectangle represents the
central part of CH.  {An AIA movie for the corresponding time
interval is available online (movie20110111.mpeg).}}
\end{figure}

\begin{figure}
\centering
\includegraphics[width=1.0\textwidth]{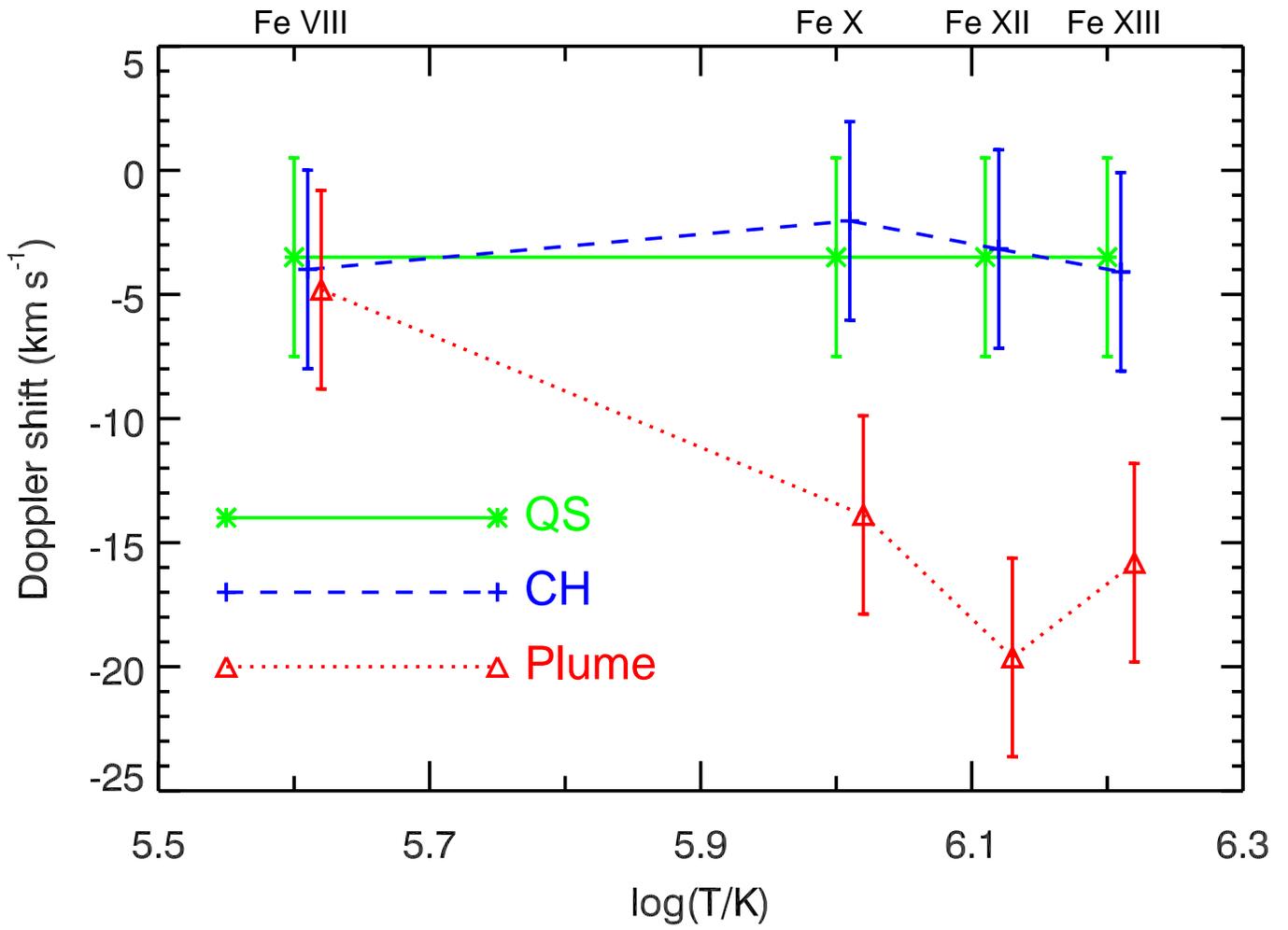}
\caption{Variation of Doppler shifts with line formation
temperature of the low latitude plume observed on 2011 Jan 11.
Green, blue and red lines represent the QS, CH and plume region,
respectively.}
\end{figure}

\begin{figure}
\centering
\includegraphics[width=1.0\textwidth]{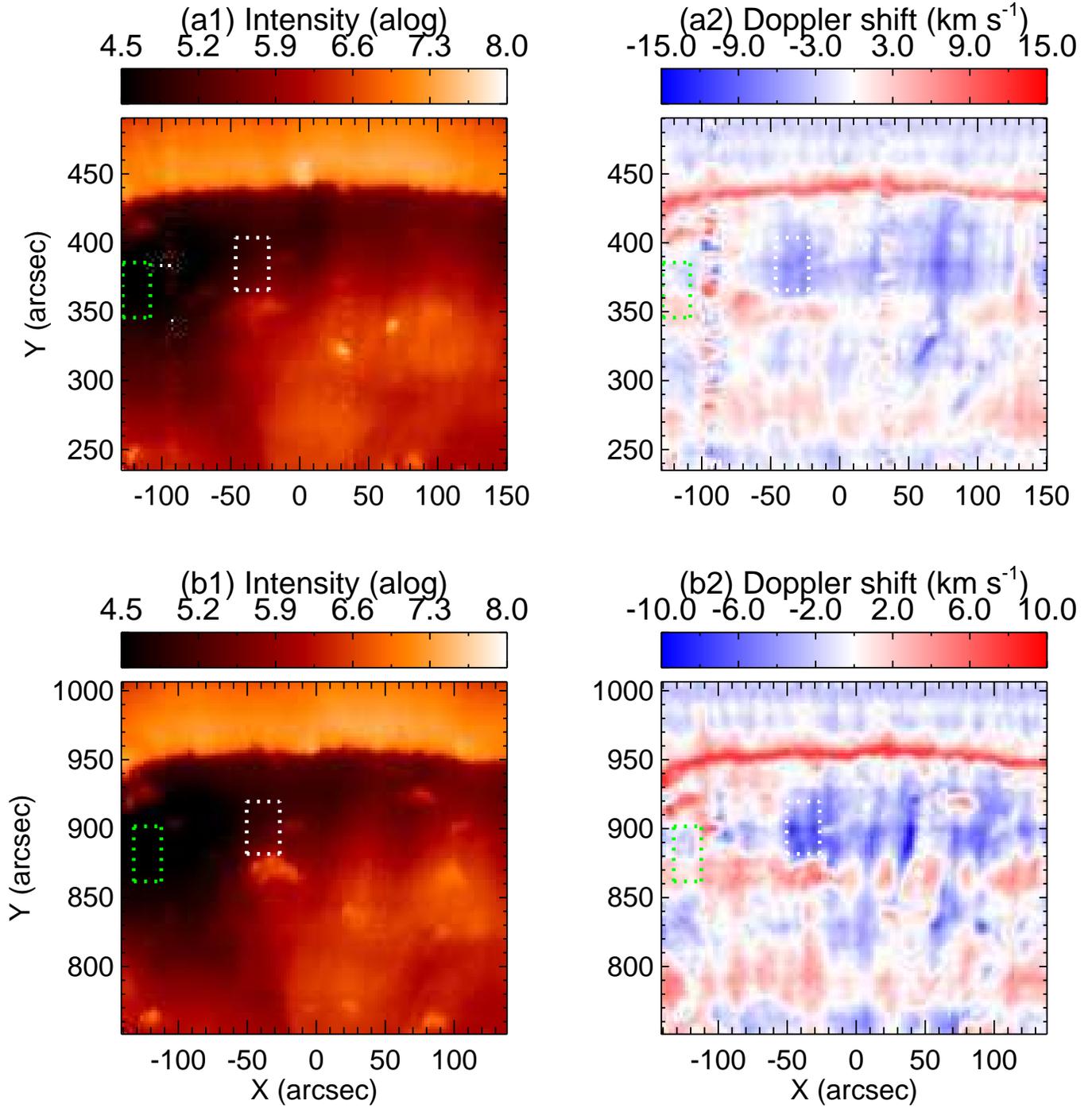}
\caption{The intensity map and Dopplergram as recorded in the
Fe~{\sc{xii}} 195.12 \AA\ line on 2010 Sep 04. The top panel was
observed from 18:30 to 22:02 and the bottom panel was observed
from 22:04 to 01:36 the following day. The selected plume is
marked by the white rectangle and the green one represents the CH
region.  {An AIA movie for the corresponding time interval is
available online (movie20100904.mpeg).}}
\end{figure}

\begin{figure}
\centering
\includegraphics[width=1.0\textwidth]{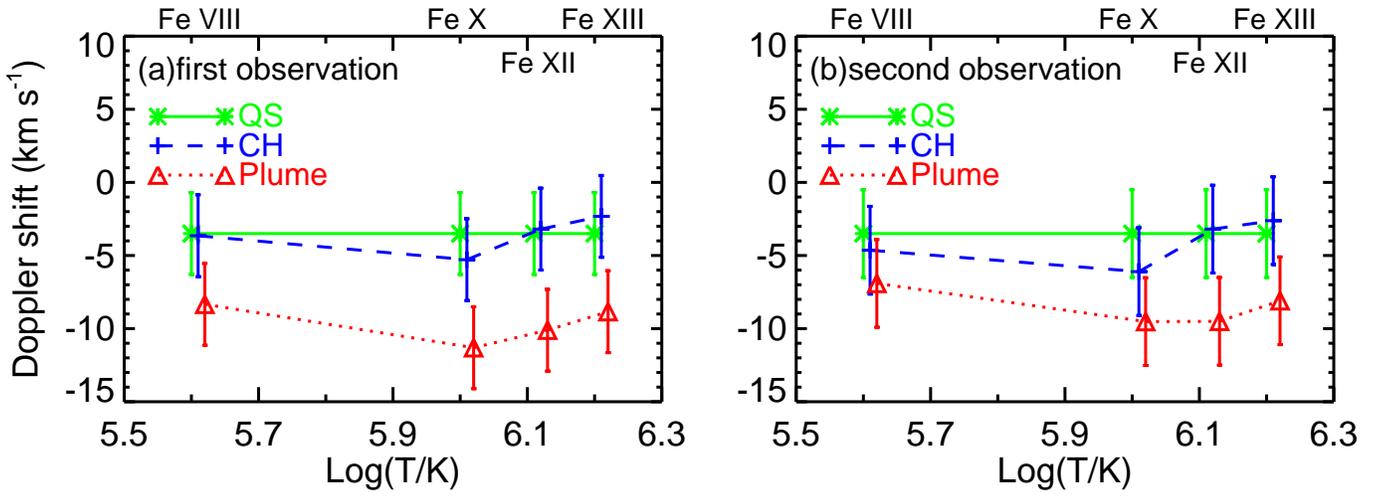}
\caption{The temperature dependence of absolute Doppler shifts in
a same plume observed 4 hours apart on 2010 Sep 04.}
\end{figure}

\begin{figure}
\centering
\includegraphics[width=1.0\textwidth]{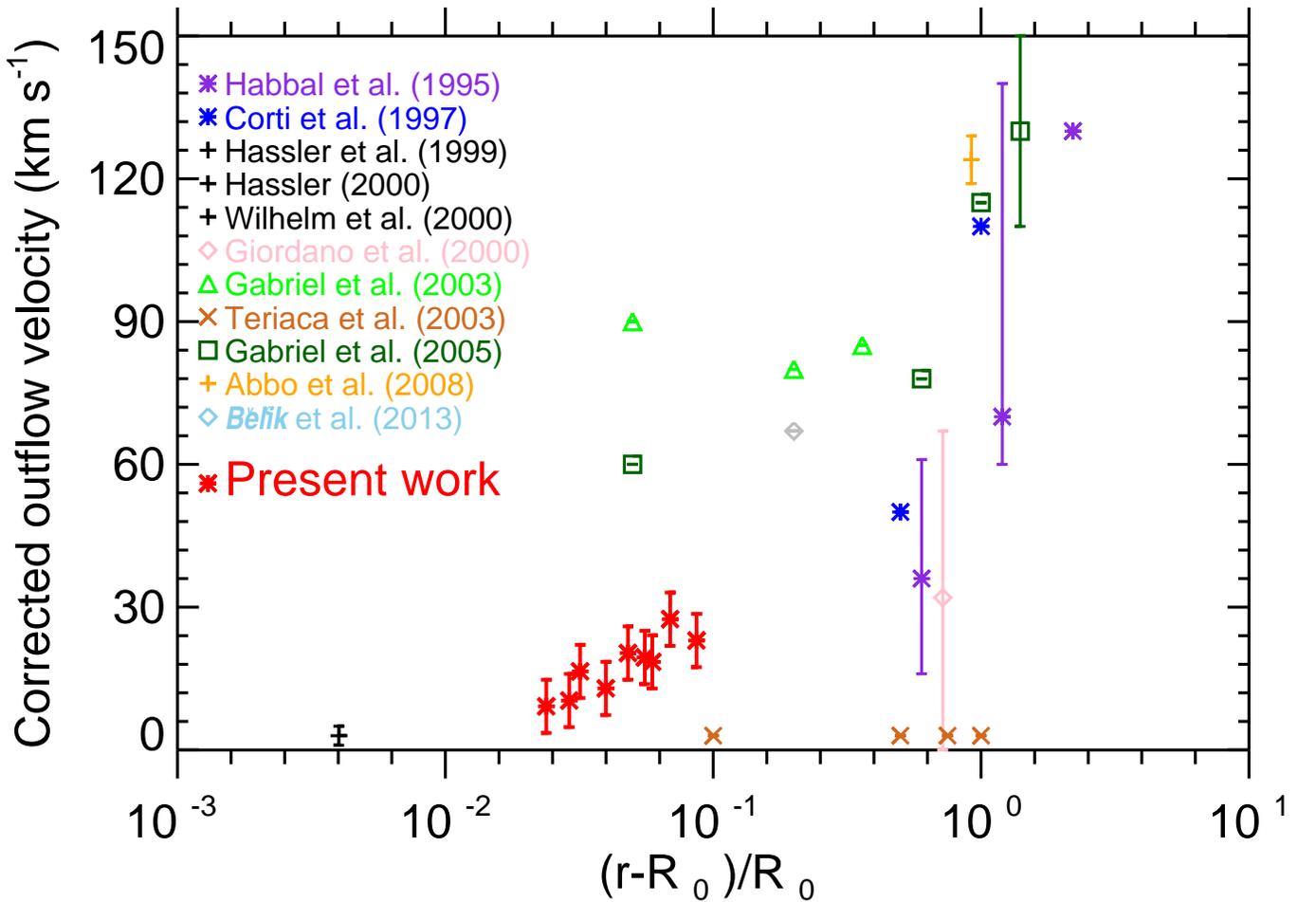}
\caption{The outflow speed in plumes at different altitudes
gathered from the literature. The results of the present work are
shown by the red asterisk.}
\end{figure}

\end{document}